\documentclass{PoS}
\let\ifpdf\relax

\usepackage{graphicx}
\usepackage{amsthm,amsmath,amssymb,bm}

\newcommand{\Tr}{{\rm Tr}}
\newcommand{\re}{{\rm Re}}
\newcommand{\im}{{\rm Im}}
\newcommand{\up}{\uparrow}
\newcommand{\down}{\downarrow}

\title{Complex Langevin simulation in condensed matter physics}

\ShortTitle{Complex Langevin simulation in condensed matter physics}

\author{\speaker{Arata~Yamamoto}\\
Department of Physics, The University of Tokyo, Tokyo 113-0033, Japan}

\author{Tomoya~Hayata\\
RIKEN Center for Emergent Matter Science (CEMS), Saitama 351-0198, Japan}

\abstract{
The complex Langevin method is one hopeful candidate to tackle the sign problem.
This method is applicable not only to QCD but also to nonrelativistic field theory, such as condensed matter physics.
We present the simulation results of a rotating Bose gas and an imbalanced Fermi-Hubbard model.
}

\FullConference{The 33rd International Symposium on Lattice Field Theory\\
                 14 -18 July  2015\\
                 Kobe International Conference Center, Kobe, Japan}

\begin{document}

\section{Complex Langevin method}
In quantum Monte Carlo simulations, ensemble average is taken on the basis of the weight of a partition function.
The weight is not always positive and can be negative or complex in the real world.
This leads to the well-known sign problem.
In lattice QCD, the sign problem is caused by chemical potentials, electric fields, theta term, and real time.
The sign problem is not a specific problem of lattice QCD.
It appears in many other quantum systems, such as matrix models, non-relativistic systems, and spin systems.
The sign problem is a crucial and broad problem in modern physics.

One hopeful way to solve the sign problem is the complex Langevin method.
In the complex Langevin method, field variables are complexified as $\Phi \in {\bf R} \to \tilde{\Phi} \in {\bf C}$.
The complex field is evolved by the complex Langevin equation
\begin{equation}
  \frac{\partial}{\partial \theta} \tilde{\Phi}
= - \frac{\delta}{\delta \tilde{\Phi}}  S[\tilde{\Phi}] + \eta ,
\label{eq:CL}
\end{equation}
where $\theta$ is a fictitious time and $\eta$ is a noise field.
The expectation value of an operator $\hat{O}$ is given by the long-time average of noise average
\begin{equation}
 \langle \hat{O} [\tilde{\Phi}] \rangle = \lim_{\theta \to \infty} \langle \hat{O} [\tilde{\Phi}(\theta)] \rangle_\eta .
\end{equation}
For the detail, see a recent review \cite{Aarts:2013uxa}.
The complex Langevin method is mainly discussed to solve the sign problem in relativistic theory \cite{Karsch:1985cb}.

In this study, we apply the complex Langevin method to two systems in condensed matter physics: a Bose system and a Fermi system.
We have two motivations.
The first one is the ab-initio analysis of condensed matter systems.
There are many unsolved systems with the sign problem.
We can analyze such systems by using the complex Langevin method. 
The second one is a test of the complex Langevin method.
Since the validity of the method has not yet understood completely, we should learn more about it.
Compared to relativistic systems, condensed matter systems are simple and the number of degrees of freedom is small.
They are good subjects for us to test the method.

\section{Bose system}

As a Bose system, we consider a nonrelativistic Bose gas \cite{Hayata:2014kra}.
The action of a complex boson field $\Phi(x) = \Phi_1(x) + i\Phi_2(x)$ is
\begin{equation}
S_0 [\Phi_1,\Phi_2] 
= \int d\tau d^3x\;\Bigl[ \Phi^{*}(x) \left( \frac{\partial}{\partial \tau}-\mu - \frac{1}{2m}\Delta \right) \Phi(x)
+\frac{1}{4}\lambda |\Phi(x)|^4 \Bigr] .
\end{equation}
In nonrelativistic systems, the imaginary-time derivative is first order and thus anti-Hermitian.
The sign problem comes from this imaginary-time derivative.
For numerical simulations, we discretize this continuum action on the (3+1)-dimensional hypercubic lattice with the lattice spacing $a$.
The lattice size is $V=N_xN_yN_zN_\tau=12^4$.
The boundary conditions are periodic.
We set $ma=0.5$ and $\lambda/a^2=4$.

\begin{figure}[h]
\centering
\includegraphics[scale=1.4]{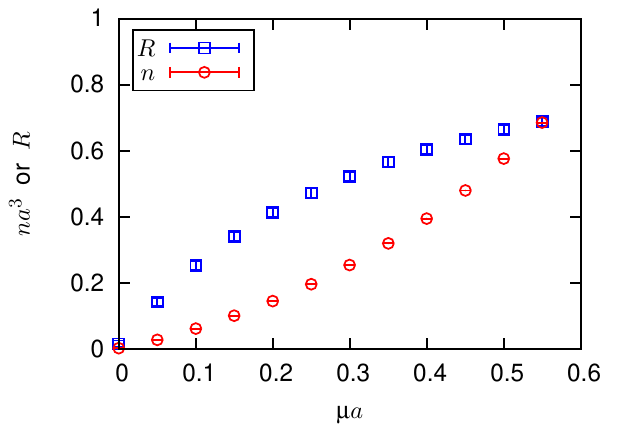}
\caption{\label{fig1}
Condensate fraction $R$ and number density $n$.
}
\end{figure}

The Bose gas becomes the Bose-Einstein condensate at low temperature.
The condensate is identified by the long-range behavior of the two-point correlation function.
When the system includes a condensate component, the condensate fraction
\begin{equation}
 R = \lim_{|x-y|\to \infty} \frac{\langle \Phi^* (x) \Phi (y) \rangle}{\langle \Phi^* (x) \Phi (x) \rangle} ,
\end{equation}
becomes nonzero.
In Fig.~\ref{fig1}, we show the simulation result of the condensate fraction and the number density.
The condensate fraction is nonzero and an increasing function of the chemical potential $\mu$.
The normal component is dominant in small $\mu$ and the condensate component is dominant in large $\mu$.

\begin{figure}[h]
 \begin{minipage}{0.5\hsize}
  \begin{center}
   \includegraphics[scale=0.35]{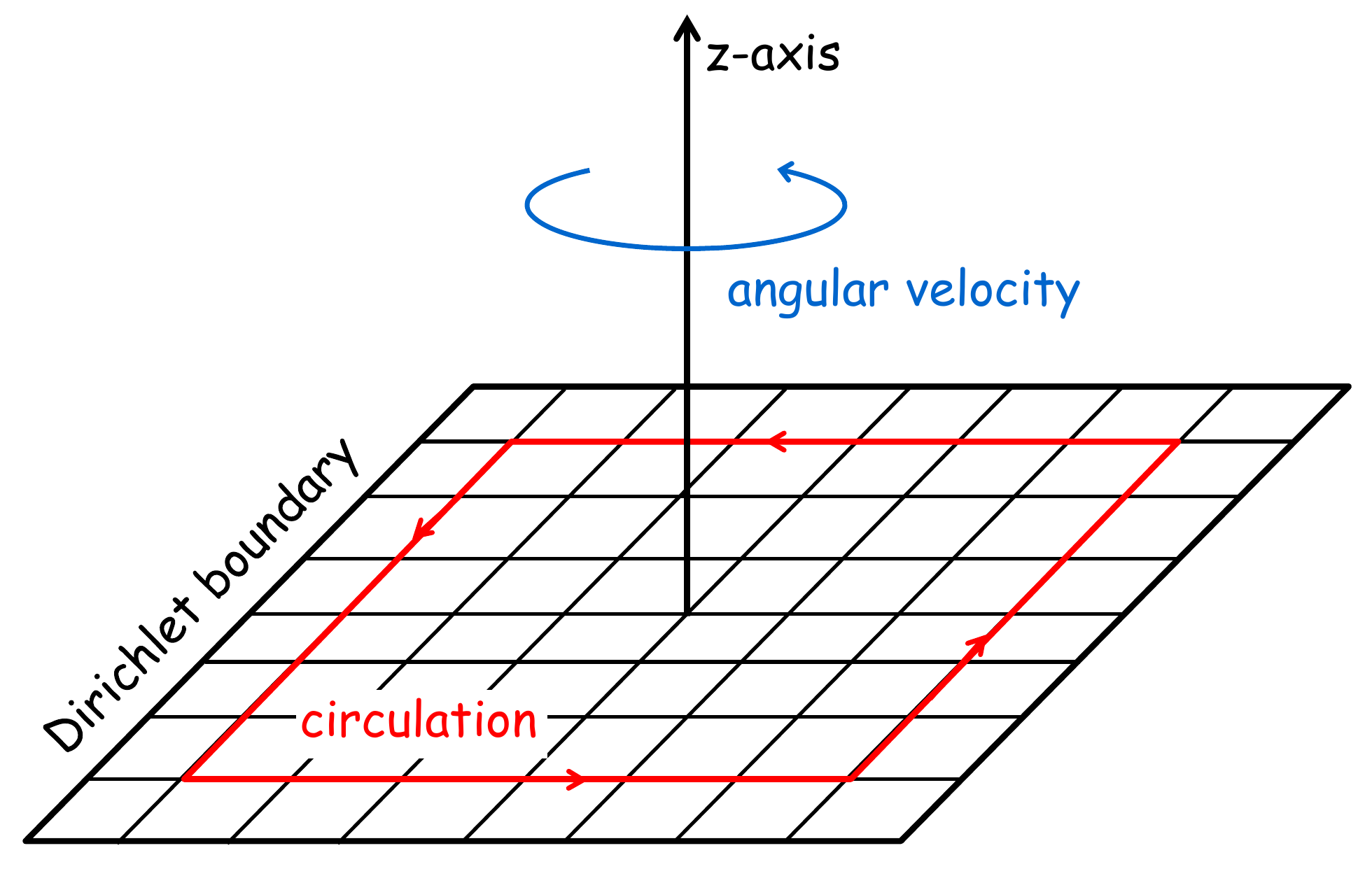}
  \end{center}
  \caption{\label{fig2}
Schematic figure of rotation.
}
 \end{minipage}
 \begin{minipage}{0.5\hsize}
  \begin{center}
   \includegraphics[scale=1.2]{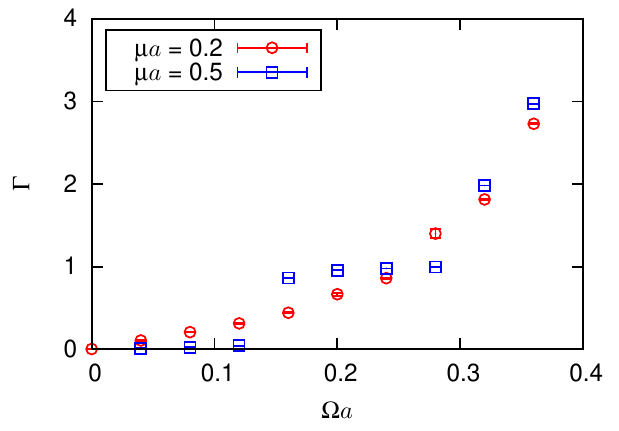}
  \end{center}
  \caption{\label{fig3}
Circulation $\Gamma$.
}
 \end{minipage}
\end{figure}

Next, we consider the rotation of this Bose-Einstein condensate.
The rotating Bose-Einstein condensate is generated and quantum vortex nucleation is observed in superfluid-helium and cold-atom experiments \cite{Fetter:2009zz}.
To describe a rotating system, we transform coordinate from a rest frame to a rotating frame.
In a rotating frame, the action shifts as
\begin{eqnarray}
S_\Omega &=& S_0 - \Omega  L_z
\\
L_z &=&  -i \int d\tau d^3x \; \Phi^{*}(x) \left( x \frac{\partial}{\partial y} - y \frac{\partial}{\partial x} \right) \Phi(x).
\end{eqnarray}
The schematic figure of the rotation is drawn in Fig.~\ref{fig2}.
Here we change the periodic boundary conditions to the Dirichlet boundary conditions in $x$ and $y$ directions.
The lattice size is $V=N_xN_y \times N_zN_\tau=11^2\times10^2$.

In the rotating Bose-Einstein condensate, quantum vortices penetrate the condensate and carry angular momentum.
The criterion for quantum vortices is the circulation.
The circulation is given by the phase integral 
\begin{equation}
\hat{\Gamma}
= \oint \frac{dx}{2\pi} \bigg[ 
  \tan^{-1} \bigg( \frac{\im[\Phi(x+\hat{j})]}{\re[\Phi(x+\hat{j})]} \bigg)
- \tan^{-1} \bigg( \frac{\im[\Phi(x)]}{\re[\Phi(x)]} \bigg) \bigg] ,
\end{equation}
along the square loop shown in Fig.~\ref{fig2}.
$\hat{j}$ is the unit vector along the loop.
The circulation $\hat{\Gamma}$ in each configuration is integer by definition.
The values fluctuate in different configurations, and thus the ensemble average $\Gamma \equiv \langle \hat{\Gamma} \rangle$ is not integer.  
In Fig.~\ref{fig3}, we show the simulation result of the circulation.
When the condensate fraction is large ($\mu a=0.5$), the expectation value of the circulation is clearly quantized.
On the other hand, when the condensate fraction is small ($\mu a=0.2$), it deviates from integer due to strong quantum fluctuation.

\section{Fermi system}

As a Fermi system, we consider the Hubbard model.
The Hubbard model is a simplified model of electrons in solids.
In particular, the two-dimensional Hubbard model is an important model of high-temperature superconductivity \cite{Scalapino:2006}. 
Also, the Hubbard model can be experimentally simulated by the optical lattice of cold atoms \cite{Esslinger:2010}.

The Hubbard model action is 
\begin{equation}
\begin{split}
S[\Psi_\up^*,\Psi_\up,\Psi_\down^*,\Psi_\down]
=& \int d\tau \sum_{x} \bigg[ \sum_{i=\up,\down} \Psi^*_i(x) \left( \frac{\partial}{\partial \tau} -\mu_i \right) \Psi_i(x)
\\
&- \sum_{i=\up,\down} \sum_{j} t_i \left( \Psi_i^*(x) \Psi_i(x+\hat{j}) + \Psi_i^*(x+\hat{j}) \Psi_i(x) \right)
\\
&+ U \Psi_\up^*(x) \Psi_\up(x) \Psi_\down^*(x) \Psi_\down(x) \bigg] ,
\end{split}
\end{equation}
where $\hat{j}$ is the unit vector to nearest neighbor sites.
For attractive interaction $U<0$, the four-fermion interaction term is rewritten by an auxiliary real scalar field $\Phi$ through the Hubbard-Stratonovich transformation.
After performing the Grassmann integral, the partition function becomes
\begin{equation}
Z = \int D\Phi \ \det K_\up[\Phi] \det K_\down[\Phi] e^{-S_A[\Phi]} = \int D\Phi \ e^{-S[\Phi]},
\end{equation}
and the total action becomes
\begin{equation}
S[\Phi] = S_A[\Phi] - \Tr \ln K_\up[\Phi] - \Tr \ln K_\down[\Phi] .
\label{eqStotal}
\end{equation}
The auxiliary field action is
\begin{equation}
S_A [\Phi] = \int d\tau \sum_{x} \frac{1}{2|U|} \Phi^2 (x) ,
\end{equation}
which is positive definite.
The fermion matrix is 
\begin{equation}
K_{i} [\Phi]
= \frac{\partial}{\partial \tau} - \mu_i + \Phi(x) - \sum_{j} t_i ( T_{+j} + T_{-j}),
\end{equation}
where $T_{\pm j}$ represents the hopping term in $j$-direction.
The condition for the sign problem is quite different from the relativistic Dirac fermion.
Since all the matrix elements of $K_i[\Phi]$ are real,  the determinant $\det K_i[\Phi]$ is real but not necessarily positive.
When two spin components are degenerate, $K_\up[\Phi] = K_\down[\Phi]$, the total determinant is semi-positive definite.
On the other hand, when two spin components are not degenerate, $K_\up[\Phi] \ne K_\down[\Phi]$, the sign problem occurs.
Unlike the Dirac fermion, a chemical potential itself is harmless but the imbalance of chemical potentials causes the sign problem.

We performed the numerical simulations of the attractive Hubbard model in two dimensions.
The hopping parameter is $t_\up a = t_\down a = 0.05$ and the total chemical potential is $\mu_\up + \mu_\down = U$.
In numerical simulations, the imaginary time $\tau$ is discretized with a temporal lattice spacing $a_\tau$.
The lattice size is $V=N_xN_y \times N_\tau=10^2\times20$.
The temperature is $aT =1/N\tau = 0.05$, where the system is in a normal phase.

\begin{figure}[h]
\centering
\includegraphics[scale=1.4]{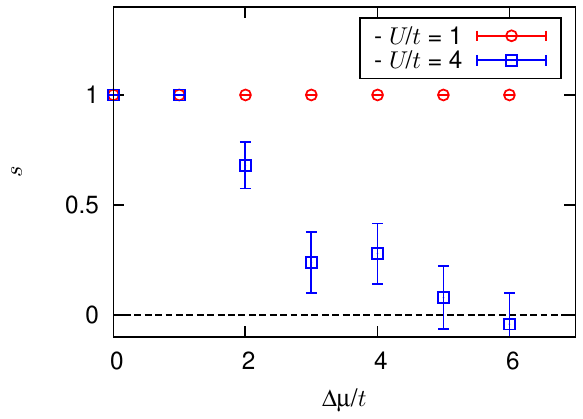}
\caption{\label{fig4}
The average sign $s$ in the sign-quenched Monte Carlo simulations.
}
\end{figure}

As a measure of the sign problem, we calculate the sign of the fermion determinant
\begin{equation}
\hat{s} = \frac{\det K_\up[\Phi] \det K_\down[\Phi]}{|\det K_\up[\Phi] \det K_\down[\Phi]|}
\end{equation}
in each configuration, and take the average in the sign-quenched Monte Carlo simulation as
\begin{equation}
s = \frac{ \int D\Phi \ \hat{s} \ |\det K_\up[\Phi] \det K_\down[\Phi]| e^{-S_A[\Phi]} }{ \int D\Phi |\det K_\up[\Phi] \det K_\down[\Phi]| e^{-S_A[\Phi]} } .
\end{equation}
This is an analogy to the average phase of the Dirac determinant in the phase-quenched lattice QCD simulation.
In Fig.~\ref{fig4}, the result is shown as a function of the chemical potential imbalance $\Delta \mu$.
At $\Delta \mu=0$, the sign is unity because of semi-positivity.
In weakly interacting case ($U/t=1$), which is far away from the phase transition, the average sign is almost completely unity.
In strongly interacting case ($U/t=4$), which is closer to the phase transition, the sign fluctuates and becomes zero for large $\Delta \mu$.
This sign fluctuation induces the sign problem.
The sign problem is more crucial when the system approaches the phase transition.

\begin{figure}[h]
 \begin{minipage}{0.5\hsize}
  \begin{center}
   \includegraphics[scale=1]{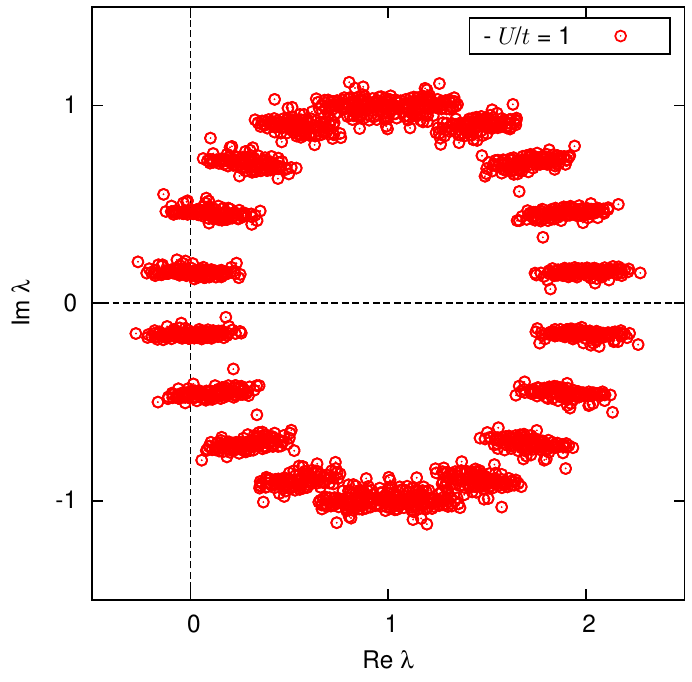}
  \end{center}
 \end{minipage}
 \begin{minipage}{0.5\hsize}
  \begin{center}
   \includegraphics[scale=1]{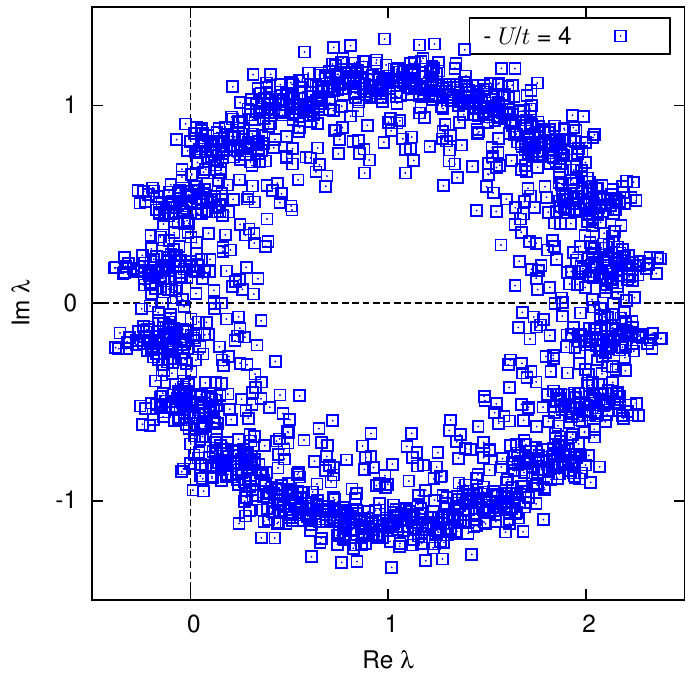}
  \end{center}
 \end{minipage}
  \caption{\label{fig5}
Eigenvalue distributions of the fermion matrix.
}
\end{figure}

In the complex Langevin method of fermion systems, we must know the eigenvalue distribution.
The eigenvalue distributions of typical configurations are shown in Fig.~\ref{fig5}.
In weak interacting case ($U/t=1$), the eigenvalue distribution has a gap around $\lambda=0$.
This gap essentially comes from the lowest fermionic Matsubara frequency $\pi T$.
In strong interacting case ($U/t=4$), the distribution spreads by the quantum fluctuation and the gap disappears.
The fermion action and the evolution equation have a singularity at zero eigenvalue.
When this singularity appears, the complex Langevin method leads a wrong result \cite{Mollgaard:2013qra}.
Therefore, the complex Langevin method fails in this fermion system because of the singularity problem.
Unfortunately, we have no idea to solve this problem at present.

In general, the singularity problem and the sign problem are separate problems.
In this particular system, however, these problems appear at the same time when the eigenvalue distribution hits $\lambda=0$.
At a high temperature or a weak coupling where is away from a phase transition, the fermion determinant is almost positive definite.
As and the system approaches the phase transition, the fermion determinant crosses zero and becomes negative.
The Monte Carlo method does not work because of the sign problem and the complex Langevin method does not work because of the singularity problem.
Only in the region away from the phase transition, the contribution of zero eigenvalues is negligible, and thus the complex Langevin method is valid.
In this region, however, the Monte Carlo method also works well.
Therefore, the complex Langevin method has no essential advantage in the view point of the sign problem.

\section*{Acknowledgments}

A.~Y.~is supported by JSPS KAKENHI Grants Number 15K17624.
The numerical simulations were carried out on SX-ACE in Osaka University.

\end{document}